# Faraday patterns in Bose–Einstein condensates

Kestutis Staliunas

*Physikalisch Technische Bundesanstalt, Bundesallee 100, D-38116 Braunschweig, Germany*

Stefano Longhi

*INFM and Dipartimento di Fisica, Politecnico di Milano, Piazza L. da Vinci 32, I-20133 Milano, Italy*

Germán J. de Valcárcel

*Departament d'Òptica, Universitat de València, Dr. Moliner 50, E-46100 Burjassot, Spain.*

## Abstract

*Temporal periodic modulation of the interatomic s-wave scattering length in Bose-Einstein condensates is shown to excite subharmonic patterns of atom density through a parametric resonance. The dominating wavelength of the spatial structures is shown to be primarily selected by the excitation frequency but also affected by the depth of the spatial modulation via a nonlinear resonance. These phenomena represent macroscopic quantum analogues of the Faraday waves excited in vertically shaken liquids.*

Spontaneous formation of spatial patterns occurs in many natural systems as well as in laboratory experiments. Some mechanisms, underlying pattern formation, such as the Turing mechanism [1] or off-resonance excitation [2], are by now well established, and some universal properties of pattern forming instabilities are understood [3]. The very essence of spontaneous pattern formation is that a uniform state loses its stability against spatially modulated states when an external control parameter is varied. The character of the instability, e.g. the dominant wavelength and symmetries of the selected patterns, is an intrinsic property of the system, independent of- (or only weakly dependent on-) initial or boundary conditions.

So far, no mechanism for spontaneous pattern formation has been suggested for the most recently created state of matter, the Bose-Einstein Condensates (BECs). Although vortices and vortex ensembles were predicted and observed in BECs [4], such structures can not be referred to as spontaneous patterns, since they occur due to particular initial or particular boundary conditions (e.g. due to stirring of the condensate). In this sense vortex ensembles are reproduced, i.e. transferred from one state of the matter into another, but do not appear spontaneously, like patterns in biological morphogenesis, Rayleigh-Benard convection, etc.

In this Letter we report that parametric excitation of a BEC can lead to spontaneous breaking of the space symmetry, and to the appearance of spatial patterns and quasipatterns. The parametric excitation is obtained by a periodic modulation in time of the scattering length of the condensed atoms. In the framework of the mean-field model of BECs (usually described by a Gross-Pitaevskii (GP) equation [5]), the temporal modulation of the scattering length of atoms corresponds to a time-dependent nonlinear coefficient. We show both by analytical and numerical calculations that the dominant wavenumber of the patterns arising from the parametric instability are selected by the excitation frequency through a dispersion-induced mechanism, and that the resulting patterns are of square, rhombic, and of eightfold symmetry, resembling the taxonomy of patterns observed in the Faraday instability on a free surface of a fluid undergoing oscillatory vertical acceleration [6]. We stress that in this work we deal with the periodic modulation of the *coefficient of nonlinearity* in the GP equation, in contrast to recent studies [7], where the periodic modulation of the *trap parameter* was investigated. In [7] it was shown that periodic modulation of the trap parameter can lead to parametric excitation of the condensate, e.g. to periodic pulsation of the condensate at the half of the excitation frequency. The modulation of the trap parameter, however, acts through the boundary conditions on the appearing structures, which are therefore rather distinct from the patterns arising from spontaneous symmetry breaking considered in this work.

We investigate the parametric excitation of patterns in condensates by considering the mean-field GP equation for the single-particle wave function $\psi(\mathbf{r},t)$ in an external trapping potential $V_{\text{trap}}(\mathbf{r})$ assuming a two-body repulsive interaction periodically modulated in time. Using suitably normalized variables the GP equation in the conservative case is [5]:

$$i\frac{\partial \psi}{\partial t} = \left[-\nabla^2 + V_{\text{trap}}(\mathbf{r}) + C(t)|\psi|^2\right]\psi, \qquad (1)$$

where $C(t) = a(t)/a_0$, $a(t)$ is the instantaneous value of the interatomic *s*-wave scattering length and $a_0$ is its average value. We consider sinusoidal modulations of the scattering length, $a(t) = a_0[1 + 2\alpha\cos(2\omega t)]$ with the frequency $2\omega$, and modulation depth $2\alpha$, hence $C(t) = 1 + 2\alpha\cos(2\omega t)$. We mainly study 2D ("pancake") BECs, with a strong trap confinement in the $z$ direction and a weak confinement, e.g. due to a harmonic trap $V_{trap}(x,y) = -1 + \frac{1}{2}\omega_{trap}^2(x^2+y^2)$ with $\omega_{trap} \ll \omega$, is the transverse directions $(x,y)$. From an experimental viewpoint, periodic modulation of the nonlinearity is possible, e.g., via the Feshbach resonance, where the variation of a magnetic field modifies the *s*-wave interatomic scattering length. This phenomenon is well known for atomic systems, and it was recently demonstrated for BECs [8]. Another possible experimental realization is to employ near-resonant laser radiation [9], or an external dc electric field [10]. The nonlinear coefficient of low dimensional (1D or 2D) condensates can also be varied by modulation of the strong confinement potential; for instance, in case of "pancake" condensates extended in the *x–y* plane the variation of trapping potential along the *z* axis can vary the resulting 2D nonlinearity [11].

We assume that the condensate is in its ground state in the absence of modulation $\psi(\mathbf{r},t) = \psi_0(\mathbf{r})e^{i\mu t}$. The wavefunction $\psi_0(\mathbf{r})$ and the chemical potential $\mu$ can be determined by a numerical analysis of Eq. (1) or by approximate analytical methods (see, e.g., [12]). Since we are interested in the occurrence of pattern forming instabilities on a spatial scale much smaller than the size of the condensate, we consider the limiting case of a flat potential and choose arbitrarily $V_{trap}(\mathbf{r}) = -1$, for which $\psi_0(\mathbf{r}) = 1$ and $\mu = 0$. We show below by direct numerical simulations that the pattern forming instability found in the flat trapping potential limit persists under a more realistic weakly-trapping harmonic potential.

In the flat trapping potential limit the spatially homogeneous solution of Eq. (1) is $\psi(\mathbf{r},t) = \psi_{hom}(t) \equiv e^{-i(\alpha/\omega)\sin(2\omega t)}$, which reduces to $\psi_0(\mathbf{r}) = 1$ in the absence of forcing. Our purpose is to determine whether a *spatially uniform*, periodic modulation of the nonlinearity coefficient is able to induce a spontaneous spatial-symmetry breaking in this homogeneous state. For that we perform a linear stability analysis of the homogeneous state against *spatially modulated* perturbations. Upon setting $\psi(\mathbf{r},t) = \psi_{hom}(t)[1 + w(t)\cos(\mathbf{k}\cdot\mathbf{r})]$, $w(t) = u(t) + iv(t)$ is the complex amplitude and $\mathbf{k}$ is the wavevector of the perturbation, and linearizing the resulting GP equation with respect to $w$, the following system of coupled equations is obtained $(k^2 = \mathbf{k}\cdot\mathbf{k})$:

$$du/dt = k^2 v, \quad dv/dt = -[k^2 + 2 + 4\alpha\cos(2\omega t)]u, \quad (2)$$

which can be combined to yield:

$$d^2u/dt^2 + [\Omega^2(k) + 4k^2\alpha\cos(2\omega t)]u = 0, \quad (3)$$

Here $\Omega(k) = k\sqrt{k^2+2}$ is the dispersion relation of the perturbations in the absence of driving. Equation (3) is a Mathieu equation which describes, e.g., a parametrically driven pendulum. Its solutions, according to Floquet's Theorem, are of the form $u(t) = \text{Re}[f(t)e^{\mu t}]$, with $f(t+\pi/\omega) = f(t)$, and the Floquet exponent $\mu = \mu(k,\omega,\alpha)$ describes the stability of the homogeneous BEC state. An instability with wavenumber $k$ arises on the neutral stability curve $\alpha = \alpha_N(k,\omega)$, implicitly defined by the relation $\text{Re}\,\mu(k,\omega,\alpha) = 0$, and the homogeneous state is unstable wherever $\text{Re}\,\mu > 0$. A general property of Eq. (3) is that its neutral stability curve is composed of an infinite series of resonance tongues located around the wavenumbers $k = k_n \equiv \sqrt{-1 + \sqrt{1+(n\omega)^2}}$, $n$ integer, which are selected by the parametric resonance condition $\Omega(k_n) = n\omega$ between the external forcing frequency $2\omega$ and the natural frequency $\Omega$ of the system, Fig. 1(a). This mechanism underlying wavenumber selection is analogous to the one generally found in parametrically forced spatially extended systems, such as in the Faraday instability of vertically shaken fluids [13].

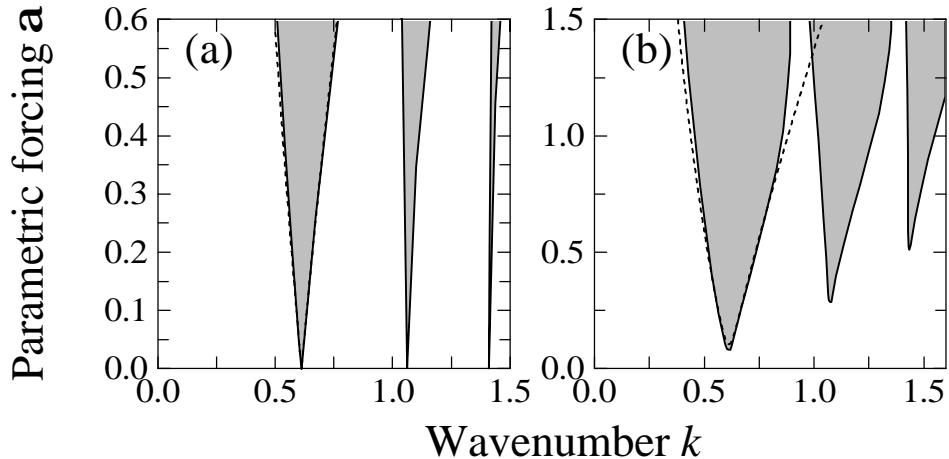

*Fig.1. Resonance tongues of the parametric instability for the conservative (a) and dissipative (b) GP equation. Shaded domains indicate the regions where the uniform condensate state is unstable. The dashed curves show the neutral stability curve for the first resonance tongue as given by Eqs.(4) and (7) in the text. Parameters values are: $\omega = 0.3 \cdot \pi$ and $\gamma = 0.03$ for Fig.1(b).*

For small values of the forcing parameter $\alpha$, an analytical expression for the neutral stability curve can be derived by a multiple scale asymptotic analysis. In particular, for the first resonance tongue $(n=1)$ it is:

$$\alpha_N(k,\omega) = \frac{\sqrt{2+k^2}}{k}|\omega - \Omega| \qquad (4)$$

Numerical simulations in the flat potential limit were performed to check the onset and evolution of the parametric instability of the homogeneous BEC state. Equation (1) was integrated using a pseudospectral split-step technique in a square spatial domain with periodic boundary conditions. Fig. 2 shows the snapshots of patterns, with the intensity of the order parameter (particle density of condensate in physical space) in the upper row, and the spatial Fourier power spectrum (particle density of condensate in momentum space) in the lower row. The figure indicates the growth of resonant modulation modes, located on concentric rings in momentum space according to the linear stability analysis. In the early stage of the development of the instability the formation of a main resonant ring in momentum space is apparent, which corresponds to transient quasipatterns in physical space, see Fig. 2(a). In the further evolution, Fig. 2(b), higher-order resonance rings appear in momentum space, which correspond to the higher-order resonance tongues of Fig. 1(a). These structures appear only transiently due to the conservative nature of the GP equation. On a long time scale one observes heating and eventual destruction of the condensate, Fig. 2(c). This destruction, which we observe in all our 2D numerics whenever the parametric instability sets in, contrasts with the everlasting periodic revivals of the spatial modulation which we observe in the 1D case.

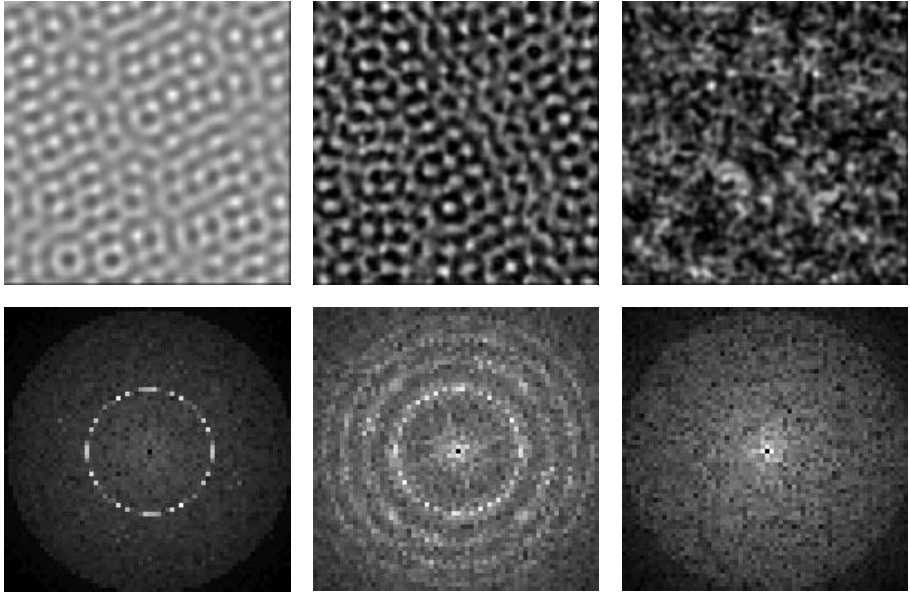

*Fig.2. Evolution of patterns in parametrically driven BECs, as obtained by numerical integration of (1) in a flat potential with periodic boundary conditions. Parameters: $a = 0.2$, $w = 1.5 \cdot p$. The snapshots were take at times: (a) $t = 100$, (b) $t = 200$, (c) $t = 300$. The zero component in momentum space pictures is removed.*

In order to determine the intrinsic symmetries of parametric patterns, we investigated weakly dissipative BECs. We expected that the inclusion of dissipative terms in the equation, capable of describing damping mechanisms of trapped BECs, can lead to the final selection of patterns with a well-defined symmetry. Theories that go beyond the GP equation providing an adequate description for the damping processes in BECs are currently the subject of intense studies. Here we adopt a phenomenological approach [14] which is compatible with experiments for dilute, alkali, Bose-condensed gas. By including damping in Eq. (1) we obtain the parametrically driven, damped GP equation:

$$i\frac{\partial \psi}{\partial t} = (1-i\gamma)\left(-\nabla^2 - 1 + |\psi|^2\right)\psi + 2\alpha\cos(2\omega t)|\psi|^2\psi. \qquad (5)$$

The damping, described by the adimensional coefficient $\gamma$, ensures an evolution towards an equilibrium state in the absence of parametric driving ($\alpha = 0$). The role of dissipation on the parametric instability is to set a non-zero threshold value for the driving strength $\alpha$ and to remove the degeneracy of threshold values for higher resonance tongues. Analogously to the derivation of Eq. (3) we find in the dissipative case the following Mathieu equation with damping:

$$d^2u/dt^2 + 2\gamma(1+k^2)du/dt + [(1+\gamma^2)\Omega^2(k) + 4k^2\alpha\cos(2\omega t)]u = 0. \qquad (6)$$

The neutral stability curves obtained from a Floquet analysis of Eq. (6) are shown in Fig. 1(b). In particular the approximate expression for the neutral stability curve of the first resonance tongue is:

$$\alpha_N(k,\omega) = \frac{\sqrt{2+k^2}}{k}\sqrt{(\omega-\Omega)^2 + \gamma^2(1+k^2)^2} \qquad (7)$$

Numerical integration of Eq. (5) with small dissipation $\gamma = 0.03$ shows the formation of stationary spatial patterns with different symmetries. Typical density distributions obtained by numerical simulations are shown in Fig. 3. In momentum space only several modes corresponding to the first resonance tongue survive due to nonlinear competition, and the pattern selection mechanism depends on the excitation frequency. For large modulation frequencies typical patterns are squares, Fig. 3(a), or quasiperiodic patterns with eightfold symmetry, Fig. 3(b). For moderate frequencies rhombic patterns are favored, Fig. 3(c). An independent validity check of the above results was performed by direct integration of a set of amplitude equations using a Galerkin decomposition, analogous to that adopted to study pattern formation in other parametrically forced systems [15]. These results also indicate patterns of square, eightfold and rhombic symmetries.

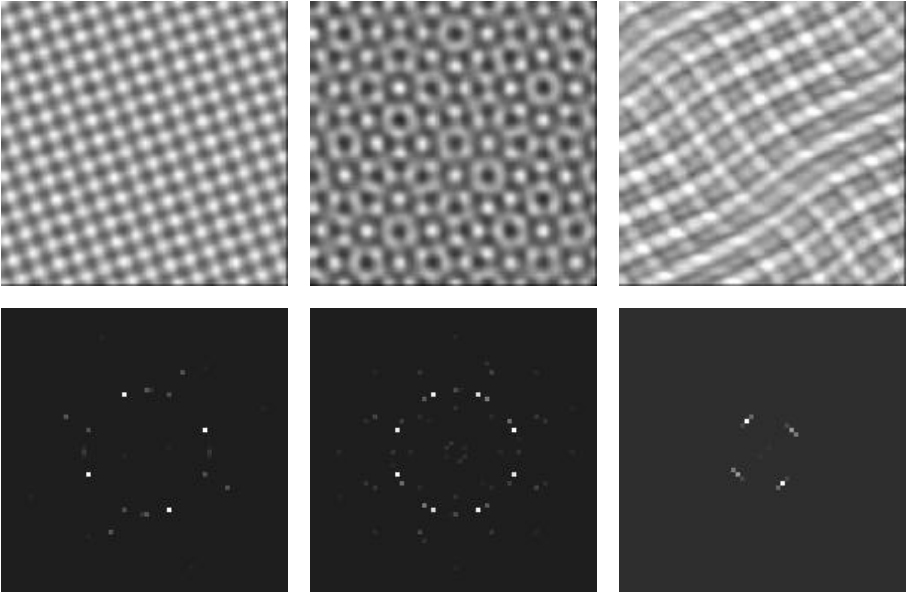

*Fig.3. Patterns in dissipative BECs, as obtained by numerical integration of (5) in a flat potential with periodic boundary conditions. Parameters are: $a = 0.2$, $g = 0.03$ and $\omega = 1.5 \cdot \pi$ for (a) and (b), and $\omega = 0.5 \cdot \pi$ for (c). The zero component in momentum space pictures is removed.*

A detailed account of the symmetries of the patterns based on a numerical study of the parametrically driven GP equation, as well as on the study of amplitude equations will be given elsewhere. We mention here that the symmetries of the observed patterns are those matching the parametric resonance condition $\Omega(k_n) = n\omega$, with $k_n \equiv \sqrt{-1 + \sqrt{1 + (n\omega)^2}}$. The scenario is the following: (i) In the linear stage of the instability a set of modes with wavevectors $\{\mathbf{k}^{(i)}, -\mathbf{k}^{(i)}\}_{i=1,2,...}$ ($|\mathbf{k}^{(i)}| = k_1$) lying on the first resonant ring are excited; (ii) Owing to nonlinear interactions, passive modes with wavevectors $\mathbf{k}^{(i,j)} = \mathbf{k}^{(i)} + \mathbf{k}^{(j)}$ are subsequently excited, with frequency $2\omega$. Among these wavevectors are some matching the parametric resonance condition associated with the second ring ($|\mathbf{k}^{(i,j)}| = k_2$). In the case of patterns composed of only two fundamental wavevectors, e.g. $\pm \mathbf{k}^{(1)}$ and $\pm \mathbf{k}^{(2)}$ with an angle $\theta$ between them, the resonance condition $|\mathbf{k}^{(1,2)}| = k_2$ is $\theta = 2\arccos(k_2/2k_1)$. For small driving frequencies ($\omega^2 \ll 1$) $\theta = \sqrt{3}\omega + O(\omega^3)$, and for large ω, $\theta = \frac{\pi}{2} + O(\frac{1}{\omega})$. These simple considerations make our observations of square patterns at large driving frequencies, and of rhombic patterns for moderate and small frequencies (Fig. 3) plausible. Also, we observe that the dominating angle of the rhombic pattern is dependent on excitation frequency. However, we were unable to ascertain definitely the above relations between the frequency of excitation and the angle. The main reason is that the local angle of the rhombic pattern usually varies in space, Fig. 3(c), also in time, thus it is not strictly defined.

The difficulties in determining the symmetries of the patterns occur possibly due to nonlinear spatial resonance. In fact we find that the selected wavenumbers depend not only on the frequency of the parametric modulation, but also on the depth of the spatial modulation of emerging pattern. Indeed, the dispersion relations are dependent on the occupation of the homogeneous condensate state $y_0$ (we normalized $y_0 = 1$ above). The resonance frequencies in the general case of the Floquet stability analysis (without using the above normalisation) are:

$$k = k_n \equiv \sqrt{-|y_0|^2 + \sqrt{|y_0|^4 + (nw)^2}} \qquad (8)$$

The growing spatial modes deplete the homogeneous state of the condensate: in the conservative case (1) the total number of the particles is conserved, therefore: $|y_0(t)|^2 = 1 - |y_{mod}(t)|^2$, where $|y_{mod}(t)|^2 = \int_{k\neq 0} |y(\mathbf{k},t)|^2 d\mathbf{k}$ is the occupation of the modulated spatial modes of the condensate. In the dissipative case (5) the total energy is conserved, therefore: $|y_0(t)|^2 = 1 - 2|y_{mod}(t)|^2$. In both cases the resonant wavenumbers are density-dependent, following (8); i.e. are always larger than the those for the undepleted condensate.

In order to give analytical evidence of this nonlinear resonance an amplitude equation for a nearly resonant, weakly nonlinear roll pattern was derived by a standard multiple scale expansion (details of the derivation are given in [16]). We expand the condensate mean field as $y(x,t) = y_{hom}(t)[1 + w(t)\cos(kx) + \sum_{n=2}^{\infty} \varepsilon^n w_n(x,t)]$, assume that the modulation depth of the pattern is small $w(t) = O(\varepsilon)$, $e$ being a small auxiliary parameter. Additionally the following parameter scalings are assumed, whose adequacy is justified a posteriory by the consistency of the final result: $\omega = \Omega(k) + O(\varepsilon^2)$ (the roll is nearly resonant), $\gamma = O(\varepsilon^2)$ (the damping is weak), and $\alpha = O(\varepsilon^2)$ (parametric driving is weak). Finally a slow time $\tau = \varepsilon^2 t$ is defined and all coefficients of the expansion are assumed to depend on both $t$ and $\tau$ [e.g., $w(t) = u(t,\tau) + iv(t,\tau)$] and $\partial_t \to \partial_t + \varepsilon^2 \partial_\tau$. Upon substituting the above expansion and scalings into Eq. (5), and equating equal powers in ε, we obtain $w(t) = (1 - \Omega/k^2)R(t)e^{i\omega t} + (1 + \Omega/k^2)R^*(t)e^{-i\omega t}$. Here the complex amplitude $R$ is governed by a Landau equation with broken phase symmetry, obtained as a solvability condition at order $\varepsilon^3$:

$$dR/dt = -(c_1 + ic_2)R + ic_3 R^* - ic_4 |R|^2 R, \qquad (9)$$

where $c_1 = \gamma(1+k^2)$, $c_2 = (\omega - \Omega)$, $c_3 = \alpha k^2/\Omega$, and $c_4 = (3+5k^2)/\Omega$. A relevant measurable quantity is the time average occupation of the nonhomogeneous modes of the condensate $r_{mod} = \langle |y_{mod}(t)|^2 \rangle$, which to the leading order is: $r_{mod} = (1 + \Omega^2/k^4)|R|^2$. The steady solution ($dR/dt = 0 : R = |R_s|e^{i\phi_s}$) of Eq. (9) results in:

$$r_{\mathrm{mod}} = \left(1 + \frac{\Omega^2}{k^4}\right)\frac{-\Omega(w-\Omega) \pm \sqrt{(ak^2)^2 - [g\Omega(1+k^2)]^2}}{3+5k^2}. \tag{10}$$

We tested this nonlinear resonance by a numerical analysis of (5) in 1D using a Galerkin expansion $y(x,t) = y_{\mathrm{hom}}(t)\sum_{n=-M}^{M} g_n(t)e^{inkx}$ with $k$ fixed and $M$ a truncation index, and calculated the time averaged intensity of the spatial modulation as a function of the excitation frequency. Fig. 4 shows the calculated nonlinear resonance: the dependence of the eigenfrequency of the spatial mode on the modulation depth is apparent, and corresponds well to that given by (10) (dashed lines). Also, bistability (subcriticality) between homogeneous and modulated solution occurs in some frequency range as the consequence of the nonlinear resonance, as Fig.4 shows.

This nonlinear resonance complicates the patterns symmetry selection since the symmetry becomes dependent not only on the driving frequency but also on the excitation amplitude through the spatial modulation depth of the condensate density. It can be expected that the nonlinear resonance affects the angle of rhombic patterns, and makes it dependent on the amplitude of the spatial modulation.

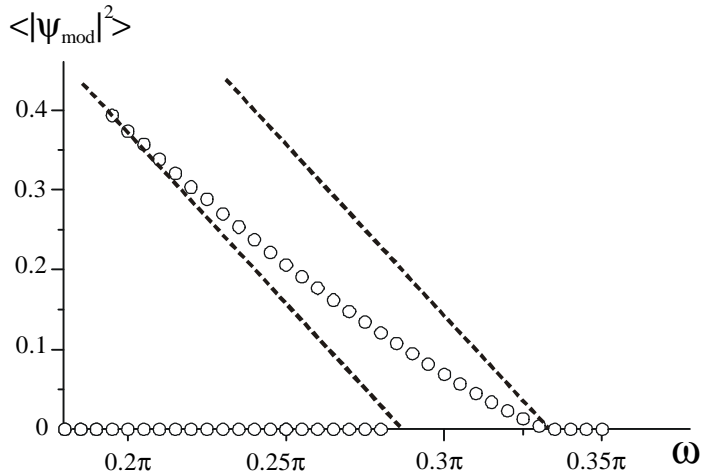

*Fig.4. Intensity of nonhomogeneous (modulated) spatial modes in dissipative BECs, as obtained by numerical integration of (5) in a Galerkin mode expansion. The set of wavenumbers of Galerkin modes is: $k_n = k_0 n$ ($n = 0, \pm 1, \pm 2,...$) where the wavelength of the lowest modulated mode $k_1 = 0.1*2p$ corresponds to frequency of $\Omega(k_1) = 0.3095p$. The dissipation parameter is $g = 0.05$, and the modulation strength is $a = 0.25$. The dashed lines are the analytic solutions of (10).*

Finally, numerical simulations were performed for the case of 2D condensates with a harmonic potential trap in order to assess the persistence of the pattern forming instability. Fig. 5 shows the spontaneous formation of patterns with a 2D harmonic potential, indicating that these structures can be observed experimentally. In physical units the calculated case corresponds to $N = 10^5$ of $^{87}Rb$ atoms in a magnetic trap of frequency $\omega_{\mathrm{trap}} = 2\pi \times 10\ s^{-1}$. This results in a condensate size (diameter at half the density maximum) equal to 50 μm with a mean scattering length $a_0 = 5.2$ nm.

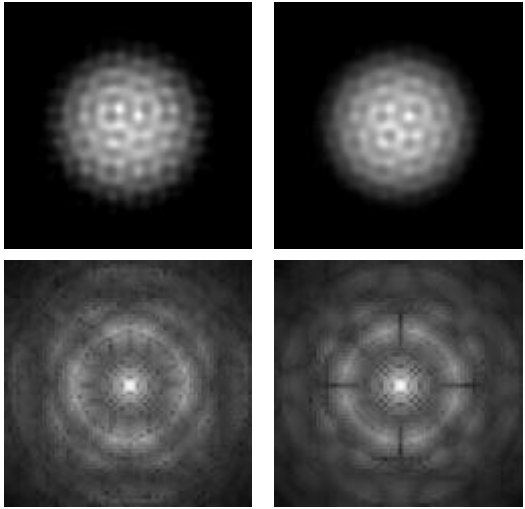

*Fig.5. Density distributions in trapped BECs, as obtained by numerical integration of (1) and (5) in a parabolic potential. Upper row - distribution in physical space, bottom row - distributions in momentum space. Left - a snapshot of transient pattern in a conservative BEC. Right- stationary pattern in a dissipative ($\Lambda = 0.03$) condensate. Parameters: $w = 1.5 \cdot p$ and $a = 0.2$ and $a = 0.4$ for conservative and dissipative case respectively. The trap frequency at these conditions was equal to $w_{trap} = 0.08 \cdot p$ for comparison.*

In conclusion, we have shown that spontaneous formation of patterns, a very general phenomenon studied in different fields of nonlinear science, is possible also for the newly created state of matter, the BECs. Pattern formation can be achieved by modulation of the scattering length of atoms in the condensate, a mechanism that bears a close connection with the formation of spatial patterns on the surface of a vibrating fluid. We studied 2D condensates, in which the dynamics occurs in the plane transverse to the tight confinement direction, and found squares, rhombi and octagons as typical patterns. Patterns with more complex symmetries are expected to occur in 3D condensates.

A nonlinear spatial resonance is also found. This nonlinear resonance makes the analysis of patterns more complicated on one hand, on the other hand it opens new possibilities for pattern formation in BECs. It is well known that a nonlinear resonance, or more precisely a subcriticality related to the nonlinear resonance, is a prerequisite for excitation of bistable spatial solitons in spatial extended systems, e.g, in nonlinear optical resonators [17]. It may be expected that such localized structures can be excited in parametrically driven BECs by employing this nonlinear resonance. We note that such localized structures, the "oscillons", are observed in periodically shaken granular materials [18], systems related to Faraday systems, and consequently related to parametrically driven BECs studied here.

The work has been supported by Sonderforschungsbereich 407 of Deutsche Forschungsgemeinschaft. G.V. acknowledges financial support from the Spanish DGES (project PB98-0935-C03-02). Discussions with M. Lewenstein, L. Santos, J. Arlt, and C. O. Weiss are gratefully acknowledged.